\documentclass[11pt]{article}
\usepackage{moriond,epsfig}

\def\Journal#1#2#3#4{{#1} {\bf #2}, #3 (#4)}

\def\SPI{\em Proc. SPIE}
\def\SPR{\em Space Sci. Rev.}

\def\be{\begin{equation}}
\def\ee{\end{equation}}
\def\bea{\begin{eqnarray}}
\def\eea{\end{eqnarray}}

\begin{document}
\vspace*{4cm}
\title{IBIS/PICsIT DATA AND ANALYSIS}

\author{J.B. Stephen \footnote{Email: stephen@tesre.bo.cnr.it}, L. Foschini \footnote{Email: foschini@tesre.bo.cnr.it},
F. Gianotti, E. Caroli, G. Malaguti, G. Di Cocco, L. Bassani, C. Ciocca, C. Labanti, A. Mauri, E. Rossi,
F. Schiavone, A. Spizzichino, A. Traci, M. Trifoglio}

\address{Istituto di Astrofisica Spaziale e Fisica Cosmica del CNR (IASF--CNR),\\
Sezione di Bologna (formerly iTeSRE), Via Gobetti 101, 40129 Bologna (Italy)}

\maketitle\abstracts{The imager IBIS on board INTEGRAL is composed of two detector
planes: ISGRI (15 keV -- 1 MeV) and PICsIT (170 keV -- 10 MeV). Here we describe
the data structure of PICsIT and the instrument specific software, including the
standard scientific analysis. Some examples of shadowgrams and spectra are shown,
by using data collected during the End--to-End Tests of January 2002.}

\section{Introduction}
The scientific objectives of the INTEGRAL satellite are the imaging and spectroscopy
of the sky in the X and gamma energy bands. The IBIS \cite{PU} telescope (15 keV -- 10 MeV)
is the imager of the satellite and consists of two detectors, ISGRI and PICsIT \cite{GM},
the former operating below about 1 MeV, while the latter functions at higher energies
(170 keV -- 10 MeV). See Table \ref{picsitcard} for a quick look at the main characteristics of IBIS/PICsIT.

To allow fine positioning of detected sources, a coded mask is situated at about
3.2 m above the PICsIT layer, so as to provide an angular resolution of 12 arcmin.
The purpose of the mask is to stop some of the incoming photons, thereby producing
"shadows" on the detector (shadowgrams are therefore the number of photons recorded
at each pixel position). To obtain the original image of the sky region, it is necessary to
perform a deconvolution of the shadowgram with the mask transfer function.
See Caroli et al. \cite{EC} for further details on coded masks.

In the following sections, the Instrument Specific Software for IBIS/PICsIT
is described, including the data structure and the scientific analysis.
Some examples of shadowgrams and spectra from the INTEGRAL End--to--End (E2E) Tests
(19 January 2001) are shown.

\begin{table}[!ht]
\caption{Short identity card for IBIS/PICsIT.}
\vspace{0.4cm}
\begin{center}
\begin{tabular}{|l|r|}
\hline
Energy range & 170 keV -- 10 MeV\\
Pixels       & 4096 ($64\times 64$) organized in 16 semimodules\\
Active area  & $\approx 2856$ cm$^2$\\
Field of View & $9^{\circ}\times 9^{\circ}$ (fully coded)\\
{}            & $19^{\circ}\times 19^{\circ}$ (half coded)\\
{}            & $29^{\circ}\times 29^{\circ}$ (zero response)\\
Angular Resolution & $12'$\\
Time Resolution &  $\approx 1$ ms\\
Point Source Location accuracy & $1'$ ($20\sigma$)\\
Energy Resolution & 12\% @ 662 keV\\
{}                & 9\% @ 1 MeV\\
\hline
\end{tabular}
\end{center}
\label{picsitcard}
\end{table}

\section{General overview}
\subsection{The IBIS/PICsIT Specific Software}
The software for IBIS/PICsIT is composed of several parts: the on board software (Hardware
Event Pre--processor of IBIS HEPI, Integral Application Software IASW) to preprocess the data before
the transmission to the ground (integration of events, multiple events reconstruction, etc.);
the \emph{Instrument Specific Software} (ISSW, that is the argument of the present paper);
the Ground Support Equipment (GSE) software; other software for modelling, MOC, ISDC, etc.

The ISSW works on the data structure created by the preprocessing of the INTEGRAL Science Data Centre
(ISDC, Geneva) and includes the Operating System Monitoring (OSM), the Automatic Calibration Analysis
(ACA), and the Standard Scientific Analysis. Once the telemetry is received on the ground (Mission
Operation Centre, MOC), it is sent to the ISDC and preprocessed. This creates the INTEGRAL
data structure, composed of raw ({\sc raw}) and prepared data ({\sc prp}). The latter consists
of the addition of the on board time to the raw data. The ISSW starts working at these levels
({\sc raw} and {\sc prp}). The OSM and ACA are part of the V1 software, while the Standard Analysis is also called V2.

The ISSW is developed by the Instrument Teams, with the help of the ISDC, which
provides some specific libraries to manage the huge data structure of INTEGRAL
(DAL). The ISSW for PICsIT is developed in C on Linux, but it is compatible also
with Sun/Solaris. In addition to the ISDC specific libraries, it makes use of standard
astronomical libraries (FITS, WCS) and format (OGIP). The software will be available
to the astronomical community through the ISDC\footnote{\texttt{http://isdc.unige.ch/}}.

The expected amount of raw data is very high: for a 100 ks observation,
PICsIT alone will produce between 1.5~Gb and 19~Gb of data, in the best and worst case
respectively, depending on the in--orbit background. The latter is a very
important unknown, able to influence the whole activity of the INTEGRAL
satellite. Only when the instrument is in the space environment,
will it be possible to perform accurate measurement of the background.

\subsection{IBIS/PICsIT modes of operation}
Because of the limited available telemetry of the INTEGRAL satellite,
the PICsIT instrument has several modes of operation.
The \emph{Standard mode} is composed of two submodes,
the \emph{spectral imaging} ({\sc spi}) and \emph{spectral timing}
({\sc spt}). In this case, the events are integrated on--board on a
characteristic time scale of 30 minutes, producing histograms.

In the spectral imaging mode, histograms are optimized for
spatial resolution, so in the integration we lose the timing
information and we reduce the energy resolution, from 1024 channels
to 256. The histograms are cubes of data with dimensions
$256\times 64\times 64$ (energy channels $\times$ pixels $\times$ pixels).
Two types of histograms are accumulated simultaneously, one for the single events, and
the other for the multiple events.

In the spectral timing mode, the purpose is to keep
the temporal resolution, so in the integration we lose the
position and the energy resolution. The histograms are divided
by default in two programmable energy bands with 1~ms time resolution.
The energy bands can be switched up to 8 by using a telecommand, and with decreasing
time resolution (down to 500 ms) to keep the same telemetry allocation.

However, in case of very low background count rate,
PICsIT can be switched to the \emph{photon--by--photon}
mode ({\sc ppm}). In this case, each event is transmitted
with its characteristic information: energy channel, now at the highest
resolution of 1024 channels, pixel position (Y, Z), and time lag between
events. The photon--by--photon mode is used also during
slews, even though the standard mode is turned on.

In all operation modes, the primary energy and incident
pixel of multiple events are reconstructed on board.

The last mode of operation of the IBIS/PICsIT instrument is the
\emph{polarimetry} mode ({\sc pol}), that will be not available
during the first year of operation. Indeed, it requires a deep
in flight calibration and analysis with real data, to understand
the efficiency and the reliability of this mode. We refer to the paper by
J.B. Stephen et al. \cite{JS} for further details.

\section{The Instrument Specific Software V1 (ISSW--V1)}

\subsection{V1: The Operating System Monitoring (OSM)}
OSM is composed of three executables (four, including {\sc pol}),
one for each mode/submode ({\sc spi}, {\sc spt}, {\sc ppm}). The purpose
of the OSM is to perform health monitoring of PICsIT.
This software works offline and therefore it is useful only for an
\emph{a posteriori} evaluation of the instrument performances. It is
not part of the Interactive OSM, developed by the ISDC for an immediate check.

The data products of the OSM are:

\begin{itemize}
\item shadowgrams in three energy bands ({\sc spi}, {\sc ppm});
\item spectra for each of the 16 semimodule ({\sc spi}, {\sc ppm});
\item lightcurves for the whole detector ({\sc spt}, {\sc ppm});
\item miscellaneous data: good pixels, counting rate statistics (mean value, standard deviation, maximum and minimum counts)
for PICsIT in all modes.
\end{itemize}

\subsection{V1: The Automatic Calibration Analysis (ACA)}
ACA is a single executable that runs once every orbit of INTEGRAL and checks the
temperatures of PICsIT semimodules from housekeeping data. If values are too much different from
reference values, the executable calculates the changes in the pixels gain and
offset and updates the tables. The relatioship between temperature and gain and offset for
each single pixel has been calculated using calibration data.

\subsection{V1: The calculation of deadtimes}
Even though this executable is part of V1, it is also used in the standard analysis (V2).
It reads livetime counters, ratemeters, and VETO strobe from housekeeping data and calculates
the deadtime for each semimodule.

\section{The Standard Scientific Analysis (ISSW--V2)}
The Standard analysis is created to produce for the final user (guest observer) a set
of scientific data, which are: images, spectra, and lightcurves. A quick summary of the
pipelines for V2, according to the IBIS/PICsIT mode of operation, is available
in Tables \ref{stapipe} and \ref{ppmpipe}. In the following sections a short
description of the intermediate tasks is shown.

\begin{table*}[!ht]
\caption{Summary of pipeline of IBIS/PICsIT V2 (Standard Mode)}
\begin{center}
\begin{tabular}{|c|c|c|c|}
\hline
Step & Images & Spectra & Lightcurves\\
{}   & {\sc spi}    & {\sc spi}     & {\sc spt}\\
\hline
1  & G\&O correction & G\&O correction & G\&O correction\\
2  & Build shadowgrams & Build shadowgrams & Rebin timing data and\\
{} & (energy bins)     & (energy bins)     & extract FOV lightcurve\\
3  & Background correction & Background correction & {}\\
4  & Deconvolution    & Single source & {}\\
{} & {}               & spectra extraction & {}\\
\hline
\end{tabular}
\end{center}
\label{stapipe}
\end{table*}

\begin{table*}[!ht]
\caption{Summary of pipeline of IBIS/PICsIT V2 (PPM Mode)}
\begin{center}
\begin{tabular}{|c|c|c|c|}
\hline
Step & Images & Spectra & Lightcurves\\
\hline
1  & G\&O correction & G\&O correction & G\&O correction\\
2  & Build shadowgrams & Build shadowgrams & Build shadowgrams\\
{} & (energy bins)     & (energy bins)     & (time bins)\\
3  & Background correction & Background correction & Background correction\\
4  & Deconvolution    & Single source & Single source\\
{} & {}               & spectra extraction & lightcurve extraction\\
\hline
\end{tabular}
\end{center}
\label{ppmpipe}
\end{table*}

\subsection{V2: Gain and offset correction}
The first step in the scientific analysis is to correct raw data for
the pixel gain and offset, and to calculate the conversion from channel to energy.
This is performed by three executables, one for each mode/submode ({\sc spi},
{\sc spt}, {\sc ppm}), which receive in input the gain and offset table,
updated by ACA every orbit.

However, although this correction is more or less easy in photon--by--photon mode,
it is not straightforward in standard mode because data are integrated in histograms.
In this case, a redistribution of photons is performed.

The correction is not applied to multiple events because they are already partially
corrected onboard by HEPI during the reconstruction of the event. Further corrections
are not possible because of the loss of some spatial information (cf also Malaguti et al.\cite{GM}).

\subsection{V2: Shadowgram building}
The following step is to build shadowgrams: in standard mode, this consists of the rebinning of the histograms,
while in photon--by--photon, the events are accumulated on the plane according to the selected energy bands.
To do this task there are two executables: one for the spectral imaging submode and the other for {\sc ppm}.
The latter is also able to build shadowgrams according to time bin (for lightcurve extraction).

\begin{figure}
\centering
\epsfig{figure=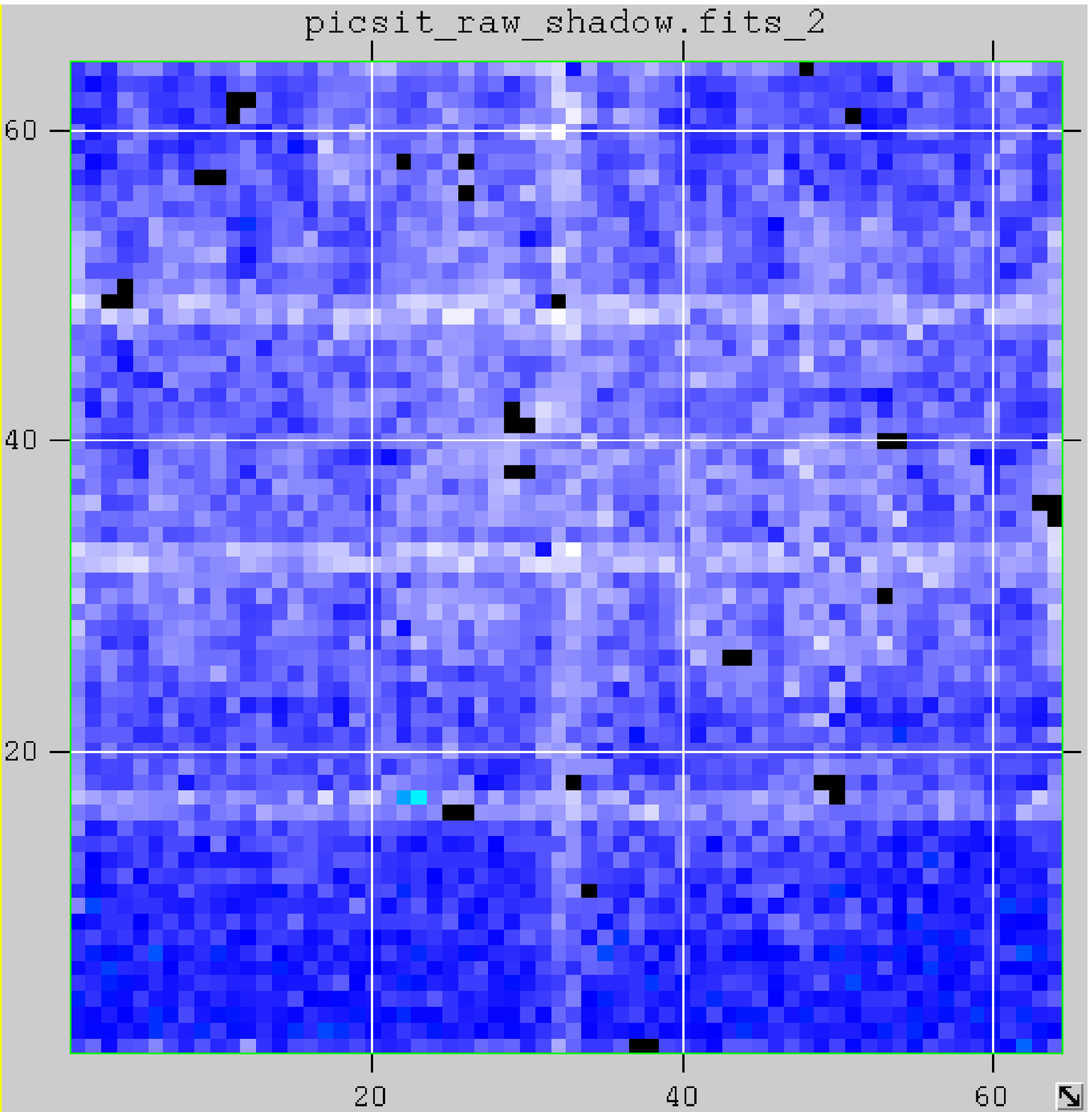, width=5cm}
\hskip 12pt
\epsfig{figure=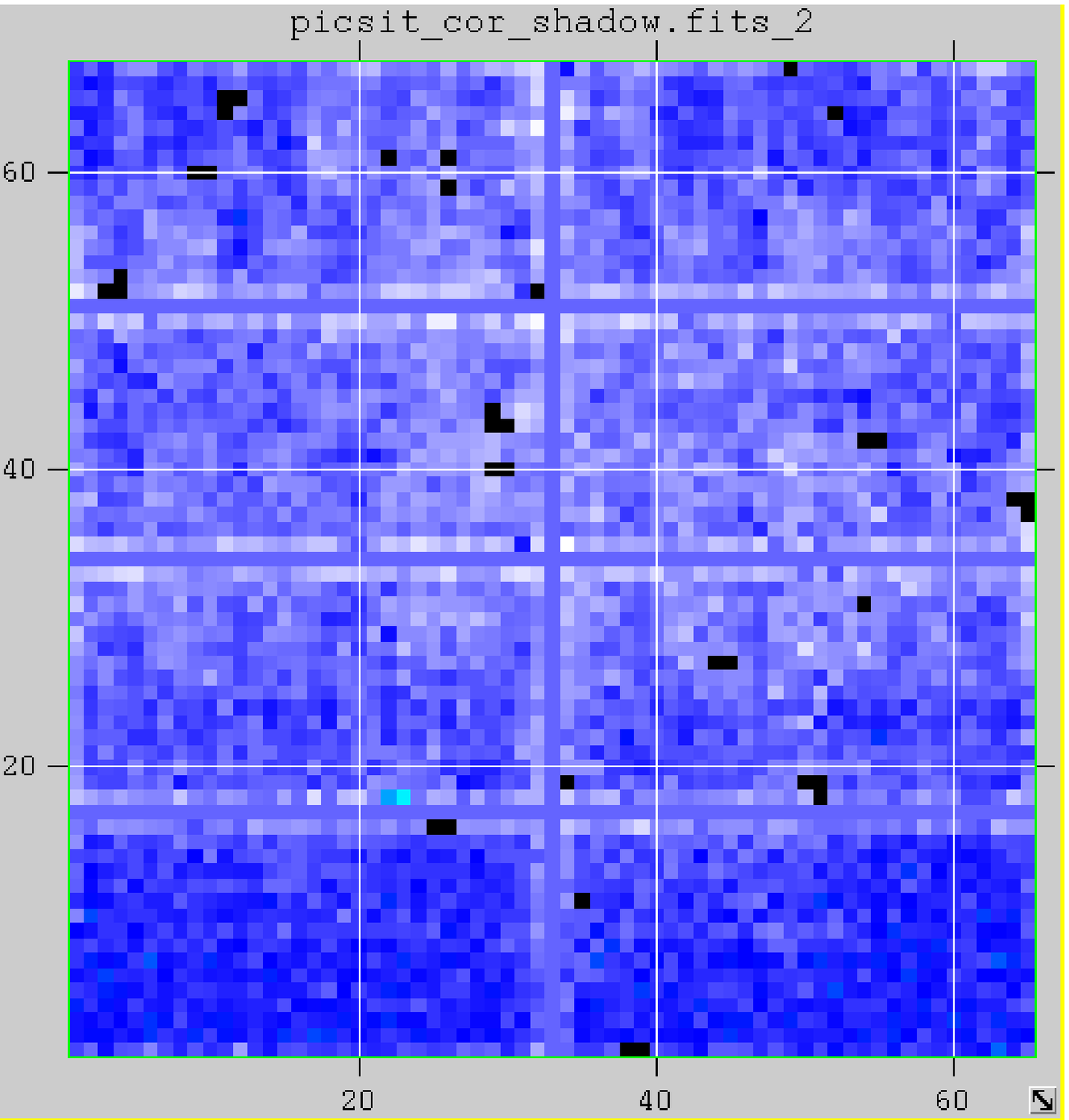, width=5cm}
\caption{Example of shadowgrams built from data of INTEGRAL E2E Tests (19 Jan 2002). (A, left) raw shadowgram; (B, right) expanded
shadowgram \label{fig:shd}}
\end{figure}

At this step, the corrections for deadtimes and faulty pixels are included. For an example of shadowgram, see Fig.~\ref{fig:shd} A.

\subsection{V2: Background correction}
Once the shadowgrams are built, before performing the deconvolution for images, spectra, or lightcurves,
it is necessary to apply some corrections to take into account the background. In addition, the gaps between
modules should be filled or tagged for null weight in the following deconvolution. To do this task,
the shadowgrams are then expanded from dimension $64\times 64$ to $65\times 67$. For an example of expanded shadowgram,
see Fig.~\ref{fig:shd} B.

\subsection{V2: Images deconvolution}
\begin{figure}
\centering
\epsfig{figure=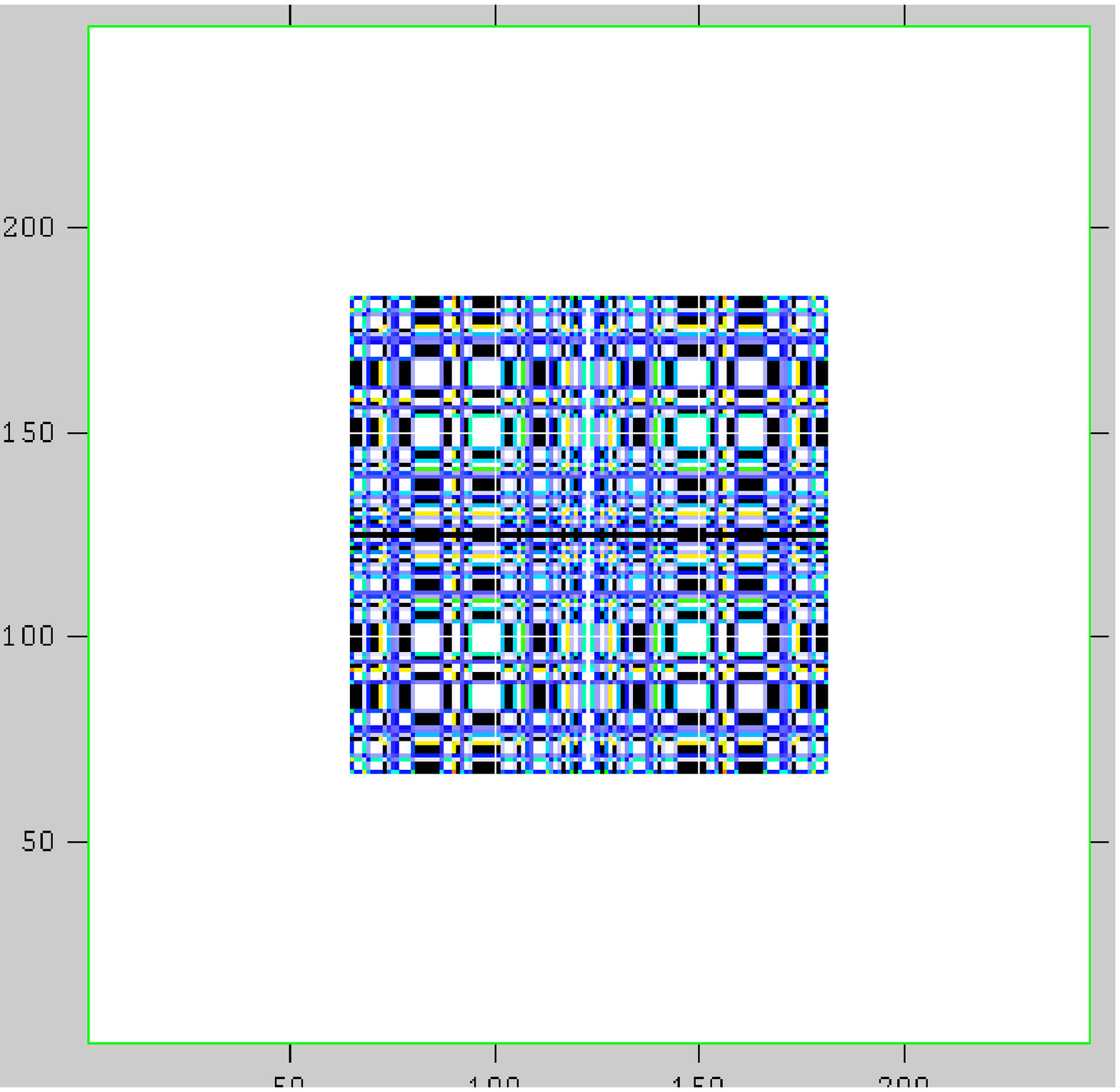, width=4cm}
\hskip 12pt
\epsfig{figure=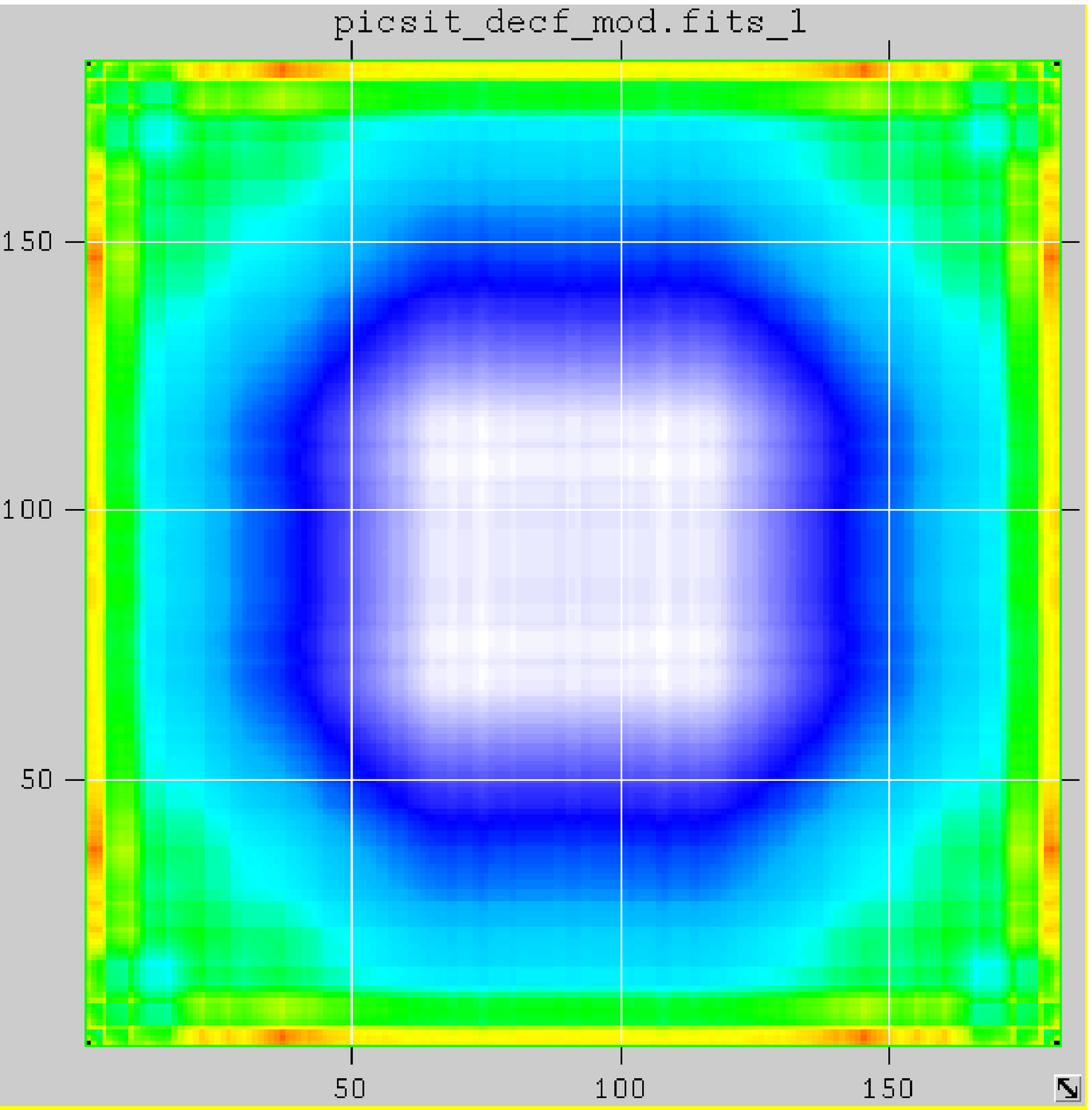, width=4cm}
\hskip 12pt
\epsfig{figure=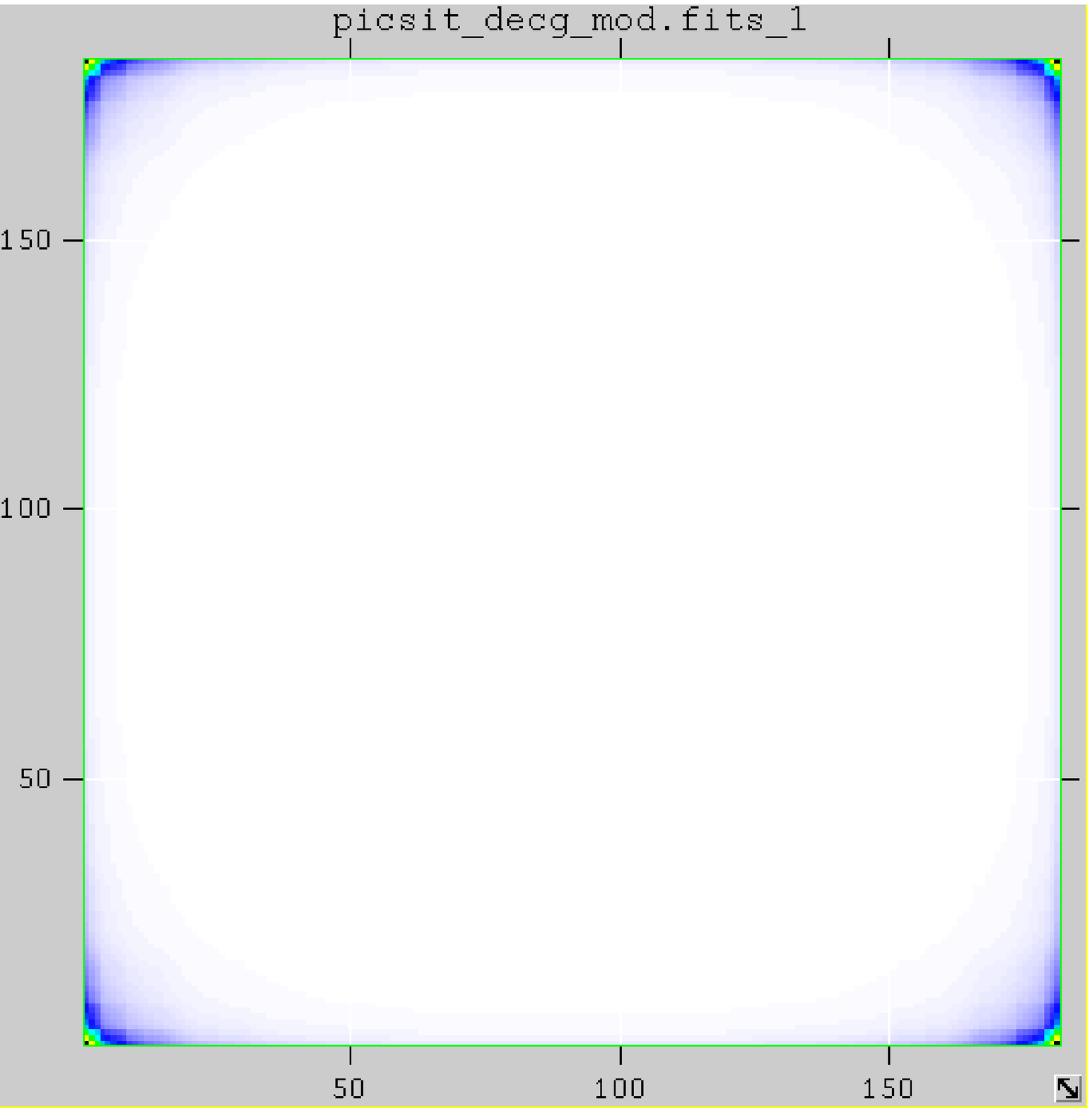, width=4cm}
\caption{Arrays used for the deconvolution: (A, left) decoding pattern; (B, centre) weighting factor F; (C, left) weighting factor G
\label{fig:deco}}
\end{figure}

IBIS detection planes are ``shadowed'' by the coded mask (from which the word ``shadowgram''
to refer to the array of counts along the detector plane). To reconstruct the sky image is then necessary
to perform a mathematical operation (deconvolution, cf Caroli et al. \cite{EC} for more details).
Specifically, the PICsIT software performs a weighted cross--correlation. The expanded shadowgram,
which is the output of the previous task, is cross--correlated with a specific decoding pattern,
which in turn is weighted according to the illuminated fraction of the mask with two other specific arrays (see Fig.~\ref{fig:deco}).

The raw image contains some byproducts of the deconvolution (ghosts, which are fake sources) and
therefore a cleaning algorithm is needed. Then, once the true sources are identified, the coordinates
and fluxes are calculated.

\subsection{V2: Spectra extraction}
The coded mask technique implies the use of non--conventional methods for the extraction
of the spectra. In fact, the photons arriving from a source in the sky are spread all over
the detector and it is possible to extract the spectrum for all the sources in the FOV
and not for a single source only.

This executable extracts all the known sources in the selected FOV and calculates how each source
illuminates the mask. Then it sums all the models of illumination and fit the results
to the data (shadowgram), checking with the $\chi^2$ test.

\subsection{V2: Lightcurve extraction}
The extraction of the lightcurve for each source is in principle the same as for extracting the spectrum.
The only difference is that the shadowgrams are now rebinned in time, instead of energy. This is if
PICsIT is in {\sc ppm} mode.

If PICsIT is in standard mode, this is not possible because the presence of the histograms.
In this case, it is only possible to extract a cumulative lightcurve for the whole FOV from
the spectral timing data. The method of using {\sc spt} data is useful if
there is a very bright source in the FOV or when there are source with known timing behaviour.

\begin{figure}
\centering
\epsfig{figure=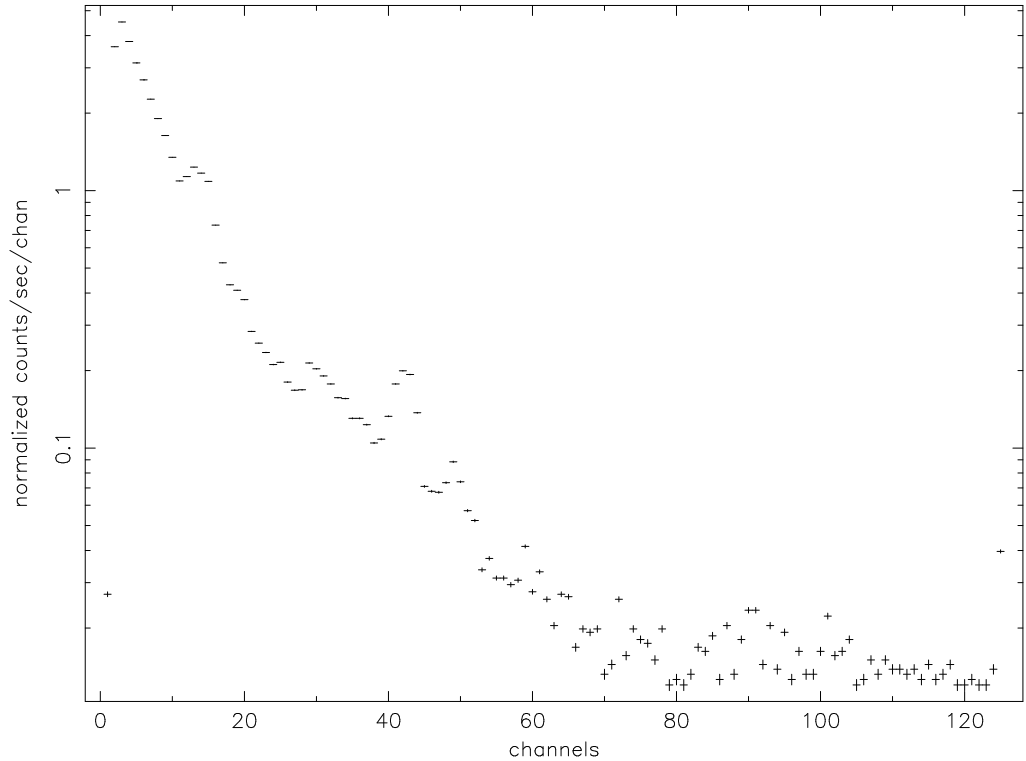, width=10cm, angle=270}
\caption{Example of PICsIT spectrum taken during the INTEGRAL E2E Tests (19 Jan 2002).
\label{fig:spex}}
\end{figure}

\section*{Acknowledgments}
We acknowledge the Italian Space Agency (ASI) for partially supporting this work.

\end{document}